\documentclass[conference]{IEEEtran}

\usepackage{graphicx}
\usepackage{xcolor}
\usepackage{hyperref}
\usepackage{setspace}
\usepackage{amsfonts}
\usepackage{mathtools}
\usepackage{algpseudocode}
\usepackage{cite}
\usepackage{amsmath,amssymb,amsfonts}
\usepackage{textcomp}
\usepackage[linesnumbered,ruled,vlined]{algorithm2e}
\usepackage{nomencl}
\usepackage{multirow}

\def\BibTeX{{\rm B\kern-.05em{\sc i\kern-.025em b}\kern-.08em
    T\kern-.1667em\lower.7ex\hbox{E}\kern-.125emX}}

\makeatletter
\def\@IEEEpubidpullup{8\baselineskip}
\makeatother

\usepackage{fancyhdr}
\usepackage{kantlipsum}
\fancypagestyle{plain}{
\fancyhf{}
\fancyhead[C]{Conference on \LaTeX} 

}
\usepackage{eso-pic}

\begin{document}

\AddToShipoutPictureBG*{
\AtPageUpperLeft{
\setlength\unitlength{1in}
\hspace*{\dimexpr0.5\paperwidth\relax}
\makebox(0,-0.75)[c]{\textbf{2023 IEEE/ACM International Conference on Advances in Social
Networks Analysis and Mining (ASONAM)}}}}

\IEEEoverridecommandlockouts
\IEEEpubid{
\parbox{\columnwidth}{\vspace{-4\baselineskip} Permission to make digital or hard copies of all or
part of this work for personal or classroom use is granted without fee provided that copies are not
made or distributed for profit or commercial advantage and that copies bear this notice and the full
citation on the first page. Copyrights for components of this work owned by others than ACM must
be honored. Abstracting with credit is permitted. To copy otherwise, or republish, to post on servers
or to redistribute to lists, requires prior specific permission and/or a fee. Request permissions from
\href{mailto:permissions@acm.org}{permissions@acm.org}.\hfill\vspace{-0.8\baselineskip}\\
\begin{spacing}{1.2}
\small\textit{ASONAM '23}, November 6-9, 2023, Kusadasi, Turkey \\
\copyright\space2023 Association for Computing Machinery. \\
ACM ISBN 979-8-4007-0409-3/23/11\ldots\$15.00 \\
\url{http://dx.doi.org/10.1145/3625007.3627313}
\end{spacing}
\hfill}
\hspace{0.9\columnsep}\makebox[\columnwidth]{\hfill}}
\IEEEpubidadjcol

\title{Generating insights about financial asks from Reddit posts and user interactions}

\author{\IEEEauthorblockN{Sachin Thukral}
\IEEEauthorblockA{\textit{TCS Research} \\
Delhi, India \\
sachi.2@tcs.com}
\and
\IEEEauthorblockN{Suyash Sangwan}
\IEEEauthorblockA{\textit{TCS Research} \\
Delhi, India  \\
suyashsangwan16@gmail.com
}
\and
\IEEEauthorblockN{Vipul Chauhan}
\IEEEauthorblockA{\textit{TCS Research} \\
Delhi, India  \\
vipul.chauhan5@gmail.com
}
\and
\IEEEauthorblockN{Arnab Chatterjee}
\IEEEauthorblockA{\textit{TCS Research} \\
Delhi, India \\
arnab.chatterjee4@tcs.com
}
\and
\IEEEauthorblockN{Lipika Dey}
\IEEEauthorblockA{\textit{TCS Research} \\
Delhi, India \\
lipika.dey@tcs.com
}
}

\maketitle

\newcommand{\ld}[1]{\textcolor{green}{[Lipika: #1]}}
\newcommand{\st}[1]{\textcolor{blue}{[Sachin: #1]}}
\newcommand{\ac}[1]{\textcolor{magenta}{[Arnab: #1]}}
\newcommand{\sj}[1]{\textcolor{cyan}{[Suyash: #1]}}
\newcommand{\vc}[1]{\textcolor{red}{[Vipul: #1]}}

\begin{abstract}
As an increasingly large number of people turn to platforms like Reddit, YouTube, Twitter, Instagram, etc. for financial advice, generating insights about the content generated and interactions taking place within these platforms have become a key research question. This study proposes content and interaction analysis techniques for a large repository created from social media content, where people’s interactions are centered around financial information exchange. We propose methods for content analysis that can generate human-interpretable insights using topic-centered clustering and multi-document abstractive summarization. We share details of insights generated from our experiments with a large repository of data gathered from subreddit for personal finance. We have also explored the use of ChatGPT and Vicuna for generating responses to queries and compared them with human responses. The methods proposed in this work are generic and applicable to all large social media platforms.
\end{abstract}

\begin{IEEEkeywords}
social network analysis, behavior, finance, Reddit
\end{IEEEkeywords}

\section{Introduction}
\label{sec:intro}

Several recent studies have shown that, unlike their predecessors, the millennials and the generations next are looking for financial advice on social media. Rather than friends or family, they prefer to rely on the advice of domain experts vetted by their peers. A recent article published by Forbes~\cite{Forbes} states that nearly $80\%$ of Americans in the age group of $18$ to $41$ get their financial advice from social media, among which Reddit and YouTube are the most trusted platforms. Independent surveys conducted among students and young professionals from Indonesia~\cite{yanto2021roles} and Uzbekistan~\cite{isomidinova2017determinants} also highlighted the strategic role played by social media in imparting financial literacy to this group, thereby influencing their financial acts. Given the above context, there is increasing interest in gaining deeper insights into the communities themselves. While the survey-based studies presented responses from a limited number of users, text analysis techniques can be applied to gather large-scale insights about the users, user-generated content, the nature of user participation, and the type of interactions manifesting within these communities. Business analysts and strategy planners take a keen interest in these insights as it provides them with a glimpse of the needs and psyche of their potential consumers, much beyond their customer base, and thereby provide them actionable intelligence to design new products, increase customer base as well as improve customer satisfaction.

In this paper, we have proposed methods for analyzing large social communities to derive insights about content and user interactions from them. We present a pipeline for completely automated analysis of textual content, along with the application of generative techniques to create human interpretable insights. We have also proposed a few novel measures for measuring interactions that can help in assessing the probable interest in and influence of generated content. The key contributions of the paper are summarized as follows:

1. A discovery-driven text analysis framework using topic modeling and topic-driven clustering is proposed for analyzing posts and comments from a large social media community. A novel text clustering algorithm has been  proposed in Sec.~\ref{sec:clustering}. The proposed algorithm is capable of discovering cohesive and distinctive topical clusters from large repositories of text data.  

2. We propose a few novel measures to capture user interaction patterns, which provide deeper contextual insights about how users interact with different kinds of content, in Sec.~\ref{sec:interactions}.  

3. We have created a large repository of publicly available data containing personal finance-related discussions, from a dedicated subreddit named \textit{r/personalfinance} which has around 18 million members. The repository was analyzed using the proposed pipeline  and a lot of interesting insights has been obtained. The results, presented in Sec.~\ref{sec:experiments} reveal insights about financial issues of persistent interest, emerging issues, and also the nature of human engagement with content.


The content analysis and interaction assessment measures proposed in this paper are fairly generic and can be applied to analyze exchanges over any large social media platform.   

\section {A framework for analyzing large social platforms}

The proposed social sensing framework consists of two key components whose functions and desirable qualities are defined as follows:

(a). Content sensing and insight generation module  - this module is responsible for analyzing large repositories of unstructured text in an unsupervised manner. Given that content generation happens in an unrestricted fashion, \textit{discoverability} is an important aspect of this component. Another desired feature of such a solution would be to ensure that all significant issues present in a repository have been identified.  



(b). Interaction sensing module - this module measures user participation in terms of different activities permitted on a social media platform like posting, commenting, upvoting or downvoting, sharing, etc. These measures help in computing the amount of user interest in content based on the extent of interactions taking place around it, and thereby help in quantifying its utilization. 

\subsection{Content Sensing using Topical Clustering} 
\label{topic}
We have used the Latent Dirichlet Allocation Machine Learning for Language Toolkit (LDA MALLET) for extracting topics\cite{rehurek2011gensim} from the posts. The MALLET topic model package consists of a fast and scalable implementation of Gibbs sampling, thereby making it highly efficient for analyzing large repositories. We argue that fewer documents should end up with equal probability for all topics, and hence $skewness$ of topic distribution for the posts is used to determine the optimal number of topics.  



For each post $P_i$, let $T_1, T_2, \ldots, T_k$ be the topical distribution of all $k$ topics given by LDA MALLET, and $\mu(P_i)$, $\sigma(P_i)$ and $\cal{M}(P_i)$ be their mean, standard deviation and median respectively.`Skewness' is a measure of the asymmetry of the probability distribution of a real-valued random variable around its mean, and is defined as:
\begin{align*}
     {\rm Skewness}(P_i) = \frac{3(\mu(P_i) - {\cal M}(P_i))}{\sigma(P_i)}.
\end{align*}
A positive skewness, in which the average topical distribution is greater than the median value of the distribution indicates that at least one of the topics is significantly present. A zero or negative skewness on the other hand indicates that none of the topics is significantly present in the post. For a given $k$, let $W_k$ denote the count of posts with negative skewness measures. Starting with a low positive integer value for $k$, say $2$, LDA is run repeatedly to extract $k$ topics, $Skewness (P_i)$ is computed for all posts $P_i$, $W_k$ that denotes the number of posts with negative skewness is computed and compared with earlier values of $k$. The value for $k$ at which the minimum value of $W_k$ is obtained, is accepted as the number of topics. This ensures that a minimum number of posts are left with equal assignment of topic probabilities, essentially indicating that their key topics cannot be determined correctly.

\subsubsection{A hybrid text clustering algorithm}
\label{sec:clustering}
Topic modeling provides high-level grouping of content, but to obtain deeper insights for deriving actionable intelligence, we now propose a topical clustering algorithm that groups together semantically similar posts within the same topic. This algorithm is executed only for representative posts of each topic, which, for a topic $T_j$  are chosen as those in which $P(T_j) > \tau$, where $\tau = 1/(k-\sqrt{k}).$ The threshold was determined through experimentation and found to be capable of identifying representative posts.


The proposed clustering algorithm is incremental in nature. Starting with a random data point as the first cluster center, it adds on cluster centers judiciously, one at a time, ensuring that the clusters created around each center are cohesive within themselves and sufficiently distinct from each other. 

Cluster cohesivity and distinctiveness are measured using the "Word Mover’s Distance (WMD)", which is known to capture both semantic and syntactic similarities between text documents effectively. WMD is a hyperparameter-free measure that computes the dissimilarity between two text documents as the minimum distance that the embedded words of one document need to “travel” to reach the embedded words of another document. The travel cost of one word $x_i$ to another word $x_j$ is defined by $c(i,j)$, the Euclidean distance between the embeddings of $x_i$ and $x_j$, and is computed as $c(i, j) = ||x_i - x_j||_2$. The word mover distance between two documents $A$ and $B$ is computed as a cumulative function of word frequencies and the minimum distance required for each word from the source document $A$ to a word in the target document $B$. It is computed as $min_{T \ge 0}(\sum_{i,j=1}^{n}\kappa_{ij}*c(i,j))$, where $\kappa_{ij}$ denotes how many times $x_i$ in document $A$ transforms into word $x_j$ in document $B$. 

We modify and extend the above formula to compute distances among multiple sets of documents, where each set eventually represents a cluster. Rather than using all words, we use a collection of keywords and phrases, identified using the YAKE tool~\cite{campos2020yake} to represent a set of documents. We replace the frequency of terms with the term frequency-inverse document frequency measure ($TF-IDF$), which can effectively capture the significant strings for each set. Keywords are embedded using Word2Vec.  
 
Let us assume that there are $n$ sets of documents, each denoted by $S_j$, where $j=1,\ldots,n$. Let $R$ denote the union of all keywords extracted by YAKE. For each set $S_j$, compute $TF-IDF$ score for all elements $w \in R$ as follows:
        \begin{align*}
            {\rm TF-IDF}(w,j) = tf(w,j) \times \log \frac{n}{df_w}
        \end{align*}
         where
         $tf(w,j)$ = number of occurrences of $w$ in $S_j$,\\
         $n$ = total number of sets,\\
         $df_w$ = number of sets that contain $w$.\\
Select the top $m$ keywords of $S_j$ sorted with respect to their $TF-IDF$ scores to represent the set. We have used $m = 10$ for our experiments.

For each pair of sets $S_i$ and $S_j$ the word mover distance between them is computed as  
    \begin{align*}
        WMD(S_i, S_j)=  \sum_{w=1}^m TF-IDF(w,j) \times c_w(i,j),
    \end{align*}
    where $c_w(i,j)$ =  minimum traveling cost of any keyword $w$ in $S_i$ to a keyword in $S_j$. 

The incremental clustering algorithm works as follows. Each post is converted into a single document vector using SBERT embeddings. Starting with the entire set of documents $D$, a random point is chosen to represent the first cluster center. The goal of the algorithm is to find subsequent cluster centers as far away as possible from all existing centers. The steps followed are presented and explained below:
    \begin{itemize}
    \item Let $c_1$ denote the randomly picked up first cluster center.
    \item Calculate the distance of all documents $d \in D$ from $c_1$ using the Euclidean distance between their SBERT embeddings.
    \item Select the furthest point from $c_1$ as the second cluster center $c_2$,
        \begin{align*}
        \mathbf{c_2} = \max_{d \in D} ||d - c_1||_2.
        \end{align*}
    \item Assign each document $d_i \in D$ to the cluster whose centre is closest to it, yielding two clusters $C_1$ and $C_2$. Recompute their cluster centers $z_1$ and $z_2$ as mean of all vectors assigned to the respective clusters. 
    \item Compute $\chi = WMD(C_1, C_2)$ as defined earlier.
    \item At this point, posts are divided into two clusters and $k = 2$. Repeat the following steps till $\chi$ stops increasing. 
    \item For each document $d \in D$, compute $\delta_{max}(d) = max_k(||d - z_k||_2)$.
    \item Choose document $p$ for which $\delta_{max}(p) \ge \delta_{max}(d)$ for all $d \in D$, as the new cluster center. Increase $k$ to $k+1$. 
    \item  Redistribute all documents $d_i \in D$ to the cluster whose center is closest to it, and recompute the cluster centers {$z_k$} as means of all vectors assigned to them.  
    \item Compute $\chi = \frac{1}{k}\sum_{i,j = 1, \ldots, k}(WMD(C_i, C_j))$, as the average word mover distance among all $k$ clusters. If new $\chi$ is greater than earlier $\chi$, then we can say clusters have moved further apart, hence this is better than the previous state.  
    \end{itemize}    


Cluster-level insights are generated using a summary of summaries approach. Posts are first summarized using OpenAI ChatCompletion model \textbf{gpt-3.5-turbo}. Cluster summaries are generated from these summaries using the same model. A few sample summaries are presented later along with other results in Sec.~\ref{sec:experiments}. 

\subsection {Measuring Platform Engagement}
\label{sec:interactions}

For a financial platform, while post content revolves around financial issues, products, services, questions or express needs for advice etc., comments in response to the posts made by authors other than the post-creator, captures the latter's interest and contribution in the post content. Voting also indicates user interest, though in a passive way.  The following measures capture user interest in a post : 

\textbf{Active Engagement around a post} is measured by the number of comments received by a post from other users. These could be in the form of suggestions, opinions, answers, or asking more questions. For computing the active engaged users (authors) on a post, we have taken the number of unique users that have commented on a post. Active engagement for a topic is computed by the average number of comments the topic has gotten over all of its representative posts.

\textbf{Passive Engagement around a post} is measured by the total score received by the post and its comments. For Reddit, the total score is given by the difference between upvotes and downvotes. The exact number of each is not known. Since each user can vote only once, the score provides the minimum number of users who must have engaged with the post, which are our passive-engaged users (voters). Passive engagement for a topic is computed as the average score the topic has received over all its representative posts and comments.

\textbf{Total Engagement} is obtained by adding the above two quantities. 
\begin{figure}[h]
    \centering    \includegraphics[width=0.8\linewidth]{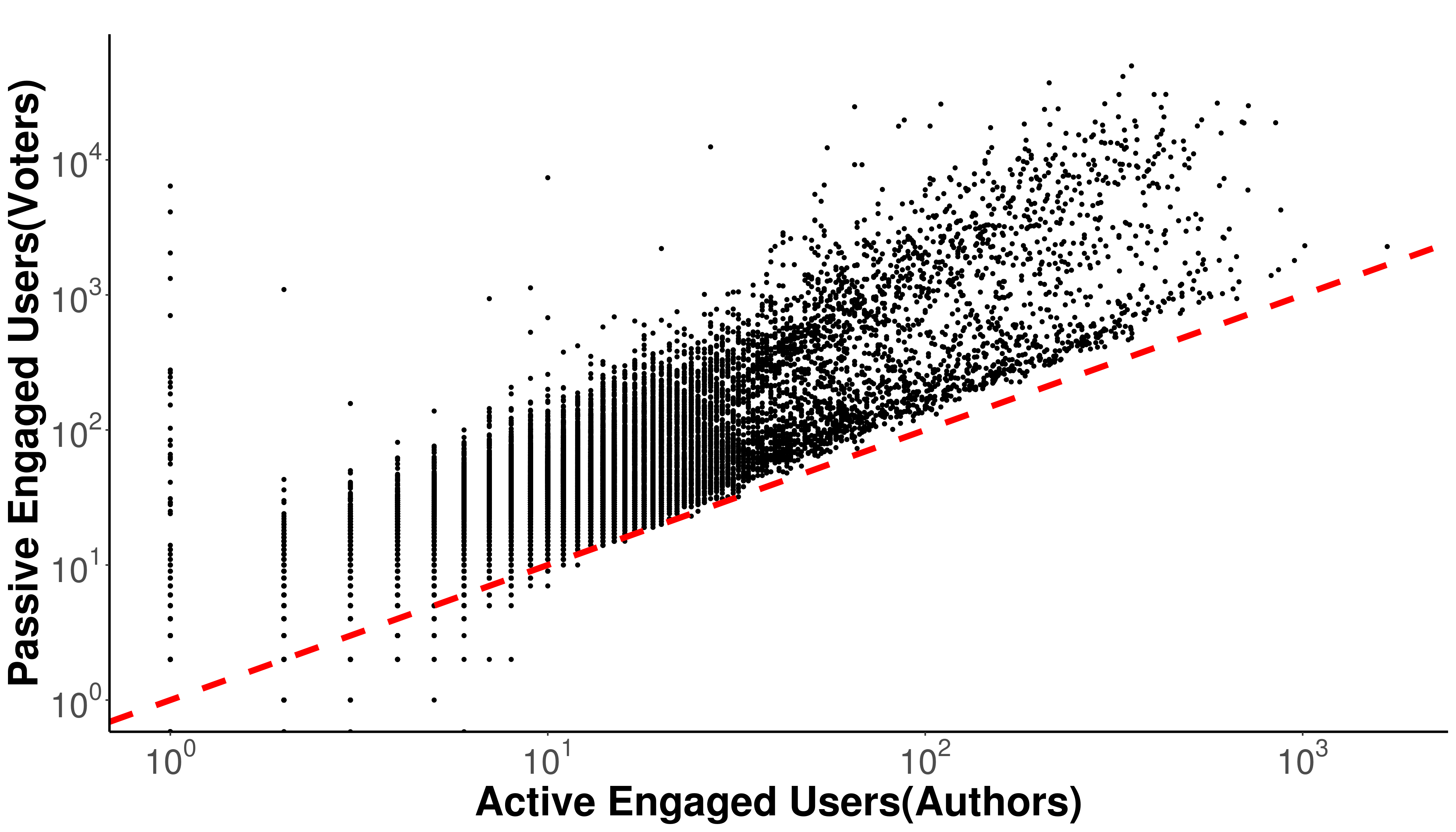}
    \caption{Engagement of Active users vs Passive users}
    \label{fig:active_total}
\end{figure}

Fig.~\ref{fig:active_total} shows a plot of active versus passive engagements computed from a financial subreddit dataset that contains $134,521$ posts and $1,521,940$ comments made by $237,718$ users. The plot shows that passive engagement for the majority of the posts is much higher than active engagement. The red dotted line indicates the equilibrium between active and passive engagements. Though the engagements have mostly garnered overall positive scores, a few dots below the red dotted line depicts posts that have garnered more downvotes than upvotes. Interestingly, their engagement levels are also very high. The vertical line of dots at the extreme left shows the number of votes received by posts, which may not have received a single comment. Manual inspection revealed that these are announcements or advertisements for financial services and products, thus indicating that people like to engage more with people than advertisers on these platforms. 

\section {Experiments, Results and Observations}
\label{sec:experiments}
 
\begin{figure}[h]
    \centering
    \includegraphics[width=0.9\linewidth]{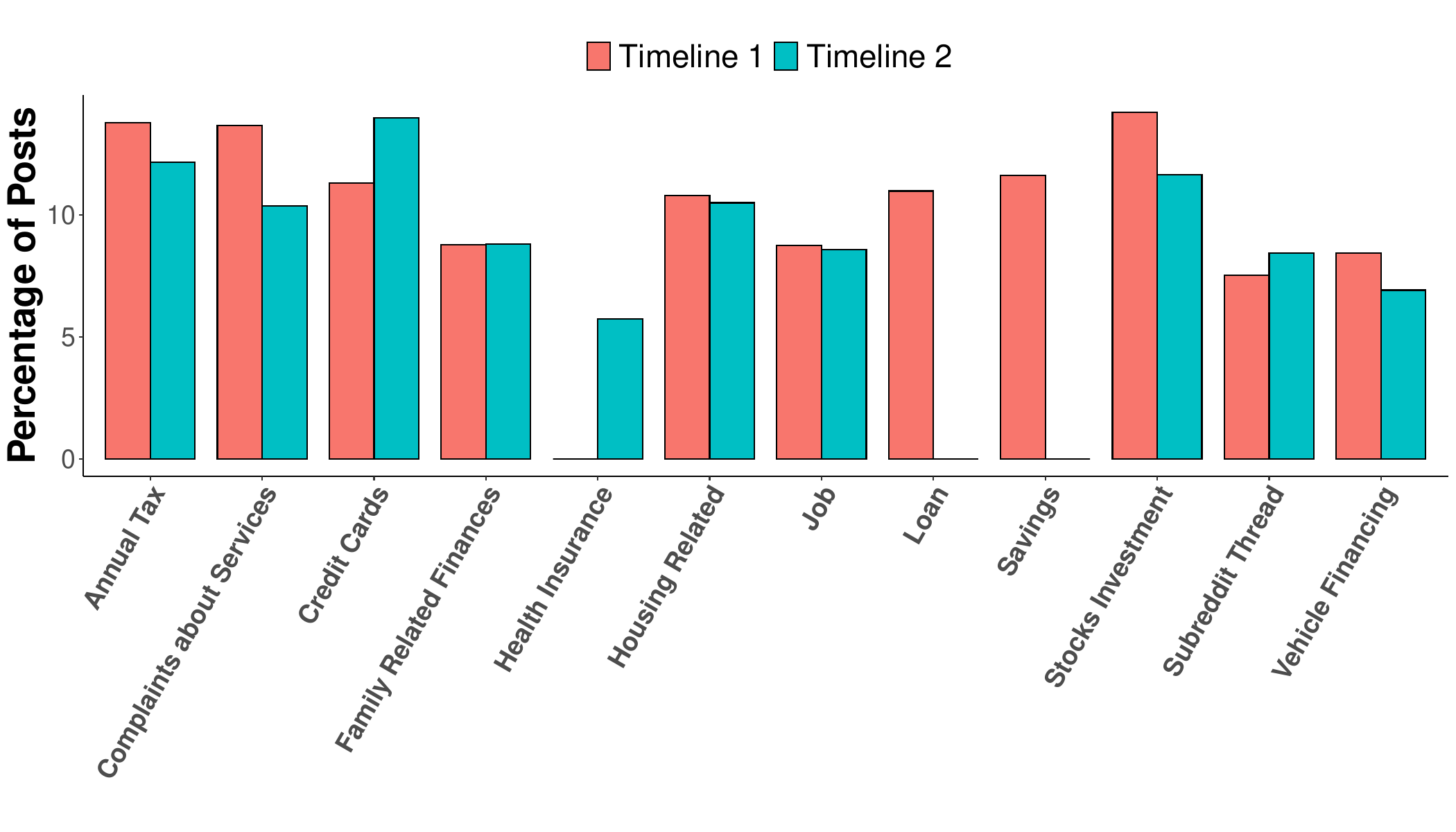}
    \caption{Topic-wise percentage of Posts in two timelines}
    \label{fig:Topicwise_percentage_of_posts}
\end{figure}

We now present insights from a large dataset which contains data for a two-year period ranging from 1st July 2020 to 30th June 2022. The data was gathered from the subreddit \textit{r/personalfinance}, using \textit{Pushshift API}~\cite{Reddit_dump}, and divided into two subsets, each of one-year duration. Table~\ref{table:dataset} presents a comprehensive account of users, posts, and comments for the two time periods. It shows an increase in the amount of content as well as number of users creating them over the period. 
\begin{table}[h]
\caption{Data set description}
	\label{table:dataset}
	\begin{center}
	    \begin{tabular}{|p{3.3cm}|r|r|r|}
	    \hline
	    Timeline & Posts & Comments &  Users \\
	    \hline
	    July 2020 - June 2021 & $134,521$ & $1,521,940$ & $237,782$ \\
            \hline
            July 2021 - June 2022 & $131,045$ & 2,041,536 & $252,572$ \\
            \hline
        \end{tabular}
	\end{center}
\end{table}
Fig.~~\ref{fig:Topicwise_percentage_of_posts} presents the topics extracted for the two timelines, along with their respective strengths. The topic names have been manually assigned based on the representative topic words returned by LDA. A number of topics like \textit{credit cards, stocks, annual tax, etc.} are present throughout. A new topic related to \textit{health insurance} has emerged in the second period, clearly due to the prevalence of the pandemic. Topics like \textit{savings} and \textit{loans} disappeared as independent topics in the second period. 

\begin{figure}[h]
    \centering
    \includegraphics[width=0.4\linewidth]{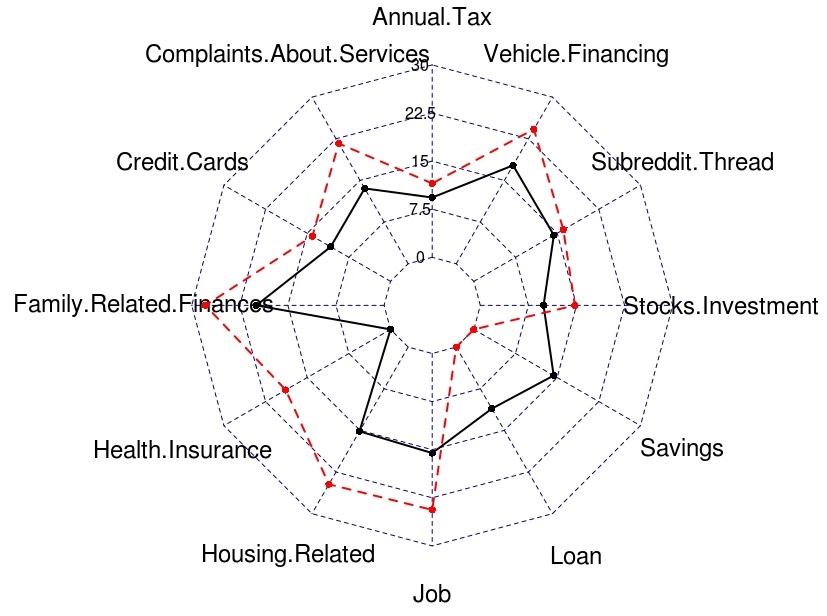}
    \includegraphics[width=0.5\linewidth]{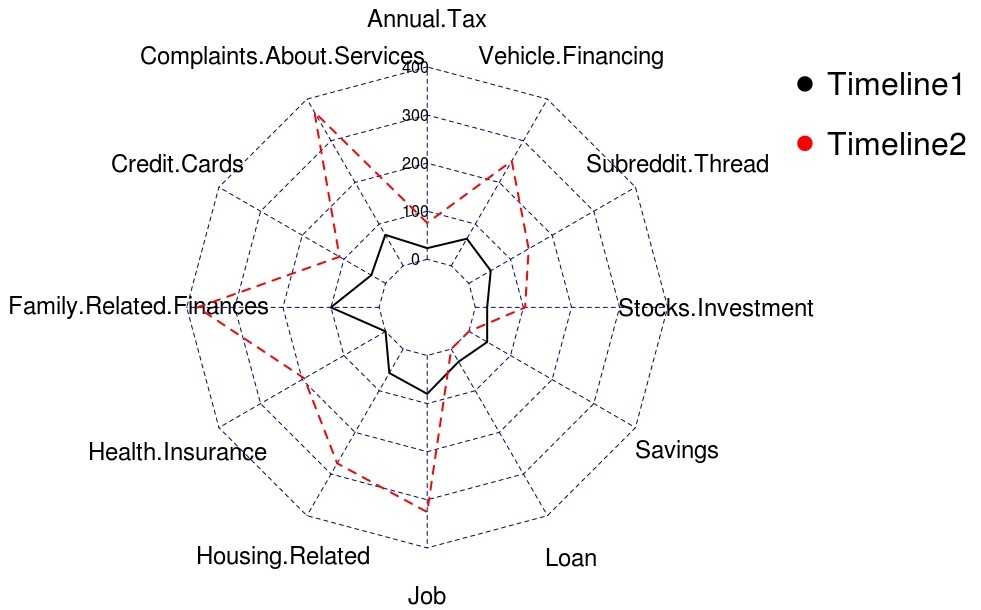}
    \caption{Topic-wise comparison of (Left) active and (Right) passive engagement in both the timelines}
    \label{fig:topic_active}
\end{figure}

Fig.~\ref{fig:topic_active} presents the levels of active and passive engagement with the different topics over the two time periods. It may be noted that while for most of the topics, both the active and passive engagement levels have increased over the years, passive engagement shows around five times increase for topics related to \textit{Family Related Finances}, and \textit{Complaints about Banking Operations}. Clearly, these platforms are helping people learn from mutual experiences. The increase in passive engagement also shows that a lot of users, who may not have shared their own queries, are following the discussions, presumably to get enriched. Thus, the platform is also playing a role in spreading awareness about managing personal finance.

On clustering the representative posts for each topic for each timeline, each topic yielded five to ten clusters. Cluster summaries generated for all using ChatGPT were also compared with an independently generated  manual summary for validation. It was found that the abstractive cluster summaries provide a succinct view of content. More details about a few key topic clusters obtained from the summaries are shared below to provide an idea about the nature of financial knowledge exchange occurring within the platform.

\textbf{Credit Cards:} User queries about credit card management clusters are around (i) strategies for improving credit scores and changing credit limits, (ii) canceling credit cards after missing payments  (iii) vehicle refinancing with a credit card, and (iv) paying medical debt with a credit card.

\textbf{Stocks and Investment:} Clusters are around (i) investment queries about Vanguard, Fidelity, and 401k options, (ii) operational issues with brokerage accounts, (iii) investment advice for retirement accounts, and (iv) seeking expert advice on portfolios and fund diversification.

\textbf{Family Related Finances:} Posts in this topic cluster are around (i) financially troubled parents, (ii) users suffering due to past family debts, (iii) seeking financial advice for mental health issues in family, and (iv) issues related to inheritance of properties and wealth. 

\textbf{Complaints about Services:} Discussions are centered around (i) issues with online refunds and purchases, (ii) identity theft, and (iii) challenges faced while closing bank accounts due to fraud. As people share their personal experiences, the discussions play an active role in creating awareness about fraud and scams. 

\textbf{Job:} These discussions are around (i) concerns about promotion and raise, (ii) seeking advice on new job offers, (iii) exploring saving options after a new job, and (iv) issues with current employer and advice sought on leaving job. 

\textbf{Health Insurance:} Discussion around health insurance schemes stood out as a unique topic during the second timeline. The clusters found are (i) seeking opinions on health plans, (ii) advice on recovering high medical bills charges, and (iii) sharing personal experiences about insurance plans.

Since the content in the platform is mostly around seeking and providing advice, we wanted to understand how a chatbot's response to these queries might compare with human responses. A small subset of the top $20$ posts from each topic of Timeline 1 were fed to ChatGPT and Vicuna~\cite{chiang2023vicuna}, and the responses were manually analyzed. The key observations about the differences in nature of human and chatbot responses are summarized in Table~\ref{table:differences}. 
 

\begin{table}[h]
\caption{Differences in Responses of Human and Machine Generated}
	\label{table:differences}
	\begin{center}
	    \begin{tabular}{|p{4cm}|p{4cm}|}
	    \hline
	    \textbf{Human Responses} & \textbf{Machine-Generated Responses} \\
	    \hline
            Large variation in length of responses & Responses are either very long or very short\\
            \hline
            Humans tend to ask questions for clarification before answering & Lacking in this ability \\
            \hline
            Ability to make decisions on user's behalf & List pros and cons and not take responsibility \\
            \hline
            May disagree with user's viewpoint & Mostly acquiescent to user's stance \\
            \hline
            More responses provide multiple viewpoints & One answer - single viewpoint \\
            \hline
            Shows empathy & Lacks Empathy \\
            \hline
            Share their own experiences & Only give to do instructions \\
            \hline
        \end{tabular}
	\end{center}
\end{table}
 

In summary, human responses on social platforms are shaped by social psychology and individual judgment. These are not based on surface-level interpretation of queries. They seek clarification and show empathy. Multiple answers provide diverse perspectives. Shared experiences connect users. Passive engagement helps in acquiring financial knowledge from these platforms. In a possible futuristic scenario, if chatbots are chosen as the only way to talk about financial needs,  the positive aspects of passive learning and obtaining diverse viewpoints from social media conversation would be lost. This may lead to biased judgments and reduced financial awareness.

\section{Related Work}
\label{sec:relatedwork}

Behavioral Finance is the study of investors’ psychology while making financial decisions. This involves analysis of market-level data, like returns and trading volume, to study the trading behavior of investors. Social media captures the activity of individuals, interactions among them, or more precisely, the complex behavior of a society.  There are studies on Twitter~\cite{oliveira2017impact} that attempt to predict returns, volatility, trading volume, and survey sentiment indices -- suggesting that mining social media may provide valuable actionable intelligence to investors. Another study suggests that differences in personalities drive decision-making in different ways~\cite{sattar2020behavioral}. Individuals interact with each other in many ways, but determining how they interact and uncovering the function of social patterns can be done using social network analysis. A recent review on social network analysis can be found in~\cite{froehlich2021social}. LDA has been used predominantly in finding the discussion topics in text data. In both of these~\cite{yin2020detecting,prabhakar2020informational}, topics were extracted using LDA from Tweets, and it was reported that LDA is able to find the most relevant and accurate topics. Another recent study focuses on finding the discussion topics in the finance domain using Reddit data~\cite{karpenko2021study}. It used RAKE to find the keywords and then distributed these keywords into topics. They have a very small data set to study for this domain as compared to ours. Recently, LLMs are also being used to study financial behavior through social media. One such study on Reddit has generated sentiment labels for the posts, using LLMs for understanding the market sentiment and being able to build a market sentiment model around it~\cite{deng2022llms}. Another study uses LLMs for financial time series forecasting on the NASDAQ stocks~\cite{yu2023temporal}. Another recent work in which users are able to develop a finance-specific LLM model named BloombergGPT, but they have said it is yet not up to the benchmark~\cite{wu2023bloomberggpt}.

\section{Conclusion} 
\label{sec:conclusion}
Users in social discussion platforms related to finance interact with each other expressing their opinions, concerns, or points of view, and also sharing their experiences. We have proposed a discovery-driven analytical framework deploying topic modeling and topic-driven clustering to derive insights about content, and also assess different types of human engagements. A large repository of public data was curated from the subreddit \textit{r/personalfinance} and insights derived from it using the proposed framework are presented. Further, we also try to assess a futuristic scenario where users may directly engage with a chatbot for answering their financial queries, rather than engage in a platform. A concern that emerges is that while social discussions involve a multitude of people and their varying points of view, one-on-one conversations with chatbots will lack multiple perspectives. Conversations may become biased in unknown ways, and its effect will be manifested on the financial activities of users. A lot of users, who are not posing questions, but following the conversation will also be affected. This might affect the level of financial knowledge in a wider population. 

One of our future tasks would be to analyze the responses in more detail and assess biases, if any. For example, we find that users often pose queries about how brands compare with each other to possibly obtain a more balanced view. Our intent is to assess whether the responses are controlled by the brands themselves. We would also like to conduct more studies to understand the demography better, for more actionable intelligence.  

\bibliographystyle{IEEEtran}
\bibliography{references}
\end{document}